# IDEAL HYDRODYNAMIC SCALING RELATIONS FOR A STAGNATED IMPLODING SPHERICAL PLASMA LINER FORMED BY AN ARRAY OF MERGING PLASMA JETS

J. T. Cassibry[1], M. Stanic[1], S. C. Hsu[2]

1. Propulsion Research Center, Technology Hall S-226 The University of Alabama in Huntsville, Huntsville, AL 35899
2. Physics Division, Los Alamos National Laboratory, Los Alamos, NM 87545

This work presents scaling relations for the peak thermal pressure and stagnation time (over which peak pressure is sustained) for an imploding spherical plasma liner formed by an array of merging plasma jets. Results were derived from three-dimensional (3D) ideal hydrodynamic simulation results obtained using the smoothed particle hydrodynamics code SPHC. The 3D results were compared to equivalent one-dimensional (1D) simulation results. It is found that peak thermal pressure scales linearly with the number of jets and initial jet density and Mach number, quadratically with initial jet radius and velocity, and inversely with the initial jet length and the square of the chamber wall radius. The stagnation time scales approximately as the initial jet length divided by the initial jet velocity. Differences between the 3D and 1D results are attributed to the inclusion of thermal transport, ionization, and perfect symmetry in the 1D simulations. A subset of the results reported here formed the initial design basis for the Plasma Liner Experiment [S. C. Hsu *et al.*, *Phys. Plasmas* **19**, 123514 (2012)].

## I. Introduction

High-velocity implosion as a mechanism for reaching high energy density (HED) states[1] is utilized, for example, in z-pinches[2], inertial confinement fusion (ICF)[3], and magneto-inertial fusion (MIF)[4,5] In the latter application, the idea of a standoff implosion driver[6–8] has motivated the Plasma Liner Experiment (PLX)[9] for exploring and demonstrating the formation of imploding spherical plasma liners using an array of merging plasma jets. This work presents the three-dimensional (3D) ideal hydrodynamic simulation results, and the scaling relations derived from them, that formed the basis for PLX, which was designed to reach ~0.1–1 Mbar of peak pressure upon the stagnation of a targetless imploding spherical plasma liner, with ~375 kJ of total initial jet kinetic energy and ~1.5 MJ of capacitive stored energy.

The liners are to be formed via the merging of thirty high Mach number plasma jets (initial n~$10^{23}$ m$^{-3}$, M~15–25, V~50 km/s, r$_{jet}$~2.5 cm) in spherically convergent geometry. Imploding spherical plasma liners could enable (1) repetitive assembly of macroscopic (cm- and µs-scale) plasmas suitable for fundamental HED physics studies and (2) further development of a standoff embodiment of MIF.[6–8]

There have been at least two related experimental research efforts within the past decade. First, the operation of a cylindrical array of 24 plasma guns[10] at up to 1 MJ of total energy has shown via fast photography that the separate gun discharges combine into a single, symmetric, cylindrically converging discharge with neutron yields up to $6\times10^9$ in pure deuterium experiments. Second, imploding solid aluminum liner experiments[11] have compressed a dense unmagnetized hydrogen plasma to Mbar pressures, as inferred from radiographs showing the radius versus time history of the target surface.

Recent work relevant to this paper include 1D radiation-hydrodynamic implosions of targetless spherical plasma liners by Awe *et al*.[12] and Davis *et al*.,[13] which showed that peak pressures achieved are underestimated unless radiation losses are included. Cassibry *et.al.*[14] conducted 3D simulations based on the ideal hydrodynamic case 6 from Table 2 of Ref. 11 and analyzed the hydrodynamic behavior of the liner during the implosion, plasma mixing during stagnation and Rayleigh-Taylor stability of the plasma throughout the entire process of jet propagation, merging, liner formation and implosion, and expansion. Kim *et.al.*[15] conducted several 1D spherically symmetric hydrodynamic simulations to explore the influence of atomic processes on the values of peak parameters, for implosions both with and without a magnetized target. More recently, Kim *et al.*[16] showed that a 3D simulation with an ionizing equation of state and initial conditions from Case 6 Table 2 of Awe et al give a peak pressure of 6.4 kbar, an order of magnitude below the corresponding spherically symmetric case. Neither the work of Cassibry *et al.*[14] nor Kim *et al.*[15,16] included radiation losses, which is one possible reason why their predicted peak pressure for case 6 of Table 2 of Ref. 11 is lower than the results of Awe *et al.*[12] and Davis *et al.*[13]. Prior computational[17] and theoretical[18] studies considered the effects of discrete jets on plasma liner implosion physics, but the results presented here constitute the first systematic study over a wide set of plasma jet parameters utilizing 3D numerical modeling.

Plasma liner implosions formed by discrete jets is a new field with no previously published peak pressure scaling results including 3D effects. While departures from ideal hydrodynamics are expected, an initial, fundamental understanding of the scaling behavior of peak pressure as a function of the multi-dimensional parameters that define the initial conditions of plasma liner formation and implosion by plasma jets is merited. Follow-on work will add and isolate non-ideal effects such as ionization and radiative/thermal conduction. This paper aims to develop a scaling relation for peak pressure as a function of the multi-dimensional initial conditions (i.e., jet velocity, Mach number, number of jets, total plasma mass, spherical jet distribution, specific heat ratio, atomic weight, and chamber radius). The purpose of finding a scaling relation is two-fold. First, such a relation provides rapid estimates of anticipated peak stagnation conditions. Second, significant departures from this scaling relation isolate the most important variables enhancing or compromising the peak pressure. For example, timing jitter and large chamber radii may significantly reduce peak pressure, while comparisons with 1D non-ideal simulations available in the open literature[12,13,15] show that the inclusion of radiation may lead to higher implosion pressures. The present simulations have been carried out using the 3D Lagrangian SPHC code,[19] with ideal gas equation of state (EOS) and without thermal conduction and radiation. A similar 3D study involving all the aforementioned physics phenomena should follow this work.

While scaling relations of the type reported in this paper have not been developed for plasma jet merging and implosion experiments, similar scaling laws have been developed for ICF. Kemp *et al.*[20] used numerical results from a self-similar solution of ideal gas dynamics of implosions of cylindrical (n=2) and spherical (n=3) shells to show that $p_s/p_0 \propto M_0^{2(n+1)/(\gamma+1)}$ for imploding shells, where $p_s, p_0,$ and $M_0$ are the stagnation pressure, maximum shell pressure, and Mach number of the imploding shell at the time of void collapse. The work by Kemp *et al.* was triggered by the observation that the minimum ignition energy scaling derived from the self-similar model is almost exactly the same as that derived from a series of 1D radiation-hydrodynamic simulations by Herrmann *et al.*[21]. Various laws have also been developed for the related fields of dense plasma focus devices[22] and z-pinches.[23] The results from Kemp *et al.* were based on a 1D self-similar model motivated by ICF physics, and do not account for the multitude of parameters in the implosion of spherical liners formed by discrete plasma jets. Accounting for such parameters is one of the primary motivations of the present study.

The remainder of the paper is organized as follows. In Sec. 2 we briefly present the main SPH principles and refer to previous work for code verification. The numerical results are given in Sec. 3, beginning with a description of the parameter space and model. The basic processes involved in plasma liner formation and implosion are then given, followed by scaling relations for merging radius and stagnation time. Since the peak pressure is determined from scattered particles as a consequence of the sph method, the interpolation method for obtaining peak pressure is then discussed. Radial plots of the pressure profile for a typical PLX-like case are given, supplemented by a 2D slice, so that the interpretation of peak pressure is clear. Peak pressure scaling was more difficult to determine than merging radius or stagnation time, and the process benefited from applying a particular scaling method. This method is presented, followed by the scaled results. Briefly, it is described by example how this relation can be used to design an experiment to produce a particular pressure. A relationship is shown between kinetic energy and peak pressure in order to estimate the total energy required for such an experiment. Conclusions are discussed in Sec. 4.

## II. Smooth particle hydrodynamics

The choice of smoothed particle hydrodynamics (SPH)[24], a free Lagrange method, for 3D plasma liner simulations was made because of difficulties in traditional Eulerian and Lagrangian algorithms using finite element or finite volume methods. Eulerian methods suffer from inaccuracies due to numerical diffusion. For 3D plasma jet applications, Eulerian grids would have to discretize a large number of nodes, of which most would consist of a vacuum, wasting computational memory and CPU time to match the resolution of a Lagrangian model in which only the jets are discretized. This is the greatest advantage of SPH, and the main reason for using it in this study.

SPH was invented by Lucy[25] and Gingold and Monaghan[26], and it has been traditionally used to investigate astrophysical processes, notably the formation of the moon and the fission of stars into binary stars[24]. SPH is a gridless Lagrangian technique[27], in which a differential interpolant of a function can be constructed from its values at the particles by using a differentiable kernel, whereby derivatives are obtained by ordinary differentiation [28]. As in finite element methods, the kernel acts as a differential test or interpolation function. For example, the integral interpolant of any function is defined by

$$A(\mathbf{r}) = \int A(\mathbf{r'}) W(\mathbf{r} - \mathbf{r'}, h) d\mathbf{r'} \quad (1)$$

where $W$ is the interpolating kernel, $r$ is the position of the particle, and $h$ is the radius of influence measured from the position $\mathbf{r}$. Numerically, Eq. 1 can be approximated by a summation interpolant

$$A(\mathbf{r}) = \sum_b m_b \frac{A_b}{\rho_b} W(\mathbf{r} - \mathbf{r_b}, h) \quad (2)$$

where $m$ and $\rho$ are the mass and density of particle $b$, respectively. Derivatives of $A$ are straightforward. For example, the gradient of $A$ is calculated as

$$\nabla A_a(\mathbf{r}) = \sum_b m_b \frac{A_b}{\rho_b} \nabla W(\mathbf{r} - \mathbf{r_b}, h) \quad (3)$$

For brevity, further discussion about SPH has been left out, but detailed theory can be found in Refs. 25 and 12. SPHC's capability to capture strong shocks has already been verified by performing simulations of the Noh problem[29,30], discussed in the previous study by Cassibry *et. al.*[14], which provides confidence in the numerical results.

**III. Numerical Simulations and Scaled Results**

For most of the cases, the variables explored were the species (distinguished by atomic mass number), number of jets $N$, initial jet velocity $V_j$, initial Mach number $M$, specific heat ratio $\gamma$, initial number density $n_j$, initial jet diameter $D_j$, initial jet length $l_j$, and initial 3D jet arrangement. Additionally, we investigated a few cases with adiabatic gas targets. We also put a slight, random perturbation in the initial radial position of each jet for a few cases, to mimic the jitter in firing multiple plasma guns simultaneously in a real experiment. All cases assumed a purely cylindrical jet with uniform properties. The effects of gradients in the jet properties and jet shape were partially considered in the study done by Stanic *et. al.*[31].

The set of initial values that cover the parameter space for scaling relations has multiple motivations. First, when exploring the influence of any parameter, it is important to examine the extreme cases within the parameter space. A good example of this in this paper is the atomic species, which varies from hydrogen to thorium. The latter was not randomly chosen as the upper limit, as thorium is applicable to fusion-fission hybrid[32,33] approaches to MIF. Most of the runs were done with argon, which is used on PLX and is a potential solution for a high-Z liner which would provide effective momentum density for magnetized target compression. As shown later, one of the most important input parameters is the plasma jet $M$, which has been varied from 1 to 100. Initial $V_j$ and jet temperature, primarily defined by the desired $M$, varied from 50 to 200 km/s and .032 to 796 eV, respectively. Since atomic processes are not modeled in this study, we utilized both high and low $\gamma$ (constant within each simulation), varying from 1.1 to 1.67 but most frequently restricted to 1.3. This way, we artificially explored the influence of the extra degrees of freedom associated with atomic processes on imploding liner evolution. Initial $n_j$ was varied over 3 orders of magnitude, ranging from $9.13 \times 10^{21}$ to $8 \times 10^{24}$ m$^{-3}$. We constrained $D_j$, $l_j$, $N$ (and their distribution), and injection radius based on what could be achieved in the near term on PLX. The number of jets modeled varied from 12 to 60, but most frequently 30 were utilized. For the 30, 36, and 60 jet cases, the spherical coverings were constrained to nodes on a truncated icosahedron (soccer ball pattern) since that is the port pattern on the PLX spherical chamber. For the 12, 18, and 24 jet cases, uniform spherical coverings were obtained from Ref. 32. All the cases are summarized in Table 1.

**Table 1. Summary of plasma jet parameters.**

| Case | Gas | N | γ | T (eV) | MW | V (km/s) | M | n (1/m³) | $D_{jet}$ (cm) | $L_{jet}$ (cm) | $r_w$ (m) |
|---|---|---|---|---|---|---|---|---|---|---|---|
| 1 | Ar | 30 | 1.3 | 2.500 | 39.948 | 50.00 | 17.8 | 1.23E+23 | 5 | 5 | 1.3716 |
| 2 | Ar | 36 | 1.3 | 2.500 | 39.948 | 50.00 | 17.8 | 1.02E+23 | 5 | 5 | 1.3716 |
| 3 | Ar | 36 | 1.3 | 2.500 | 39.948 | 50.00 | 17.8 | 1.23E+23 | 5 | 5 | 1.3716 |
| 4 | Xe | 30 | 1.3 | 2.500 | 131.293 | 50.00 | 32.4 | 3.74E+22 | 5 | 5 | 1.3716 |
| 5 | Ar | 30 | 1.3 | 2.500 | 39.948 | 50.00 | 17.8 | 3.00E+22 | 8 | 8 | 1.3716 |
| 6 | Xe | 30 | 1.3 | 2.500 | 131.293 | 50.00 | 32.4 | 9.13E+21 | 8 | 8 | 1.3716 |
| 7 | Ar | 30 | 1.3 | 2.500 | 39.948 | 70.00 | 25.0 | 1.53E+23 | 8 | 8 | 1.3716 |
| 8 | Xe | 30 | 1.3 | 2.500 | 131.293 | 70.00 | 45.3 | 4.66E+22 | 8 | 8 | 1.3716 |
| 9 | Ar | 12 | 1.3 | 2.487 | 39.948 | 200.00 | 71.6 | 5.00E+23 | 15.24 | 50 | 1.3716 |
| 10 | Ar | 18 | 1.3 | 2.487 | 39.948 | 200.00 | 71.6 | 5.00E+23 | 15.24 | 50 | 1.3716 |
| 11 | Ar | 24 | 1.3 | 2.487 | 39.948 | 200.00 | 71.6 | 5.00E+23 | 15.24 | 50 | 1.3716 |
| 12 | Ar | 24 | 1.3 | 2.487 | 39.948 | 200.00 | 71.6 | 5.00E+23 | 15.24 | 50 | 1.3716 |
| 13 | Ar | 60 | 1.3 | 2.487 | 39.948 | 200.00 | 71.6 | 5.00E+23 | 15.24 | 50 | 1.3716 |
| 14 | Ar | 24 | 1.3 | 2.487 | 39.948 | 200.00 | 71.6 | 5.00E+23 | 15.24 | 25 | 1.3716 |
| 15 | Ar | 24 | 1.3 | 2.487 | 39.948 | 200.00 | 71.6 | 5.00E+23 | 15.24 | 12.5 | 1.3716 |
| 16 | Ar | 24 | 1.3 | 2.487 | 39.948 | 50.00 | 17.9 | 8.00E+24 | 15.24 | 50 | 1.3716 |
| 17 | Ar | 24 | 1.3 | 2.487 | 39.948 | 100.00 | 35.8 | 2.00E+24 | 15.24 | 50 | 1.3716 |
| 18 | Ar | 24 | 1.3 | 2.487 | 39.948 | 150.00 | 53.7 | 8.89E+23 | 15.24 | 50 | 1.3716 |
| 19 | Ar | 24 | 1.3 | 2.487 | 39.948 | 200.00 | 71.6 | 5.00E+23 | 15.24 | 50 | 1.3716 |
| 20 | Ar | 24 | 1.3 | 2.487 | 39.948 | 200.00 | 71.6 | 5.00E+23 | 15.24 | 50 | 1.3716 |
| 21 | Ar | 24 | 1.3 | 2.487 | 39.948 | 200.00 | 71.6 | 5.00E+23 | 15.24 | 50 | 1.3716 |
| 22 | Ar | 24 | 1.3 | 2.487 | 39.948 | 200.00 | 71.6 | 5.00E+23 | 15.24 | 50 | 1.3716 |
| 23 | Ar | 24 | 1.3 | 2.487 | 39.948 | 50.00 | 17.9 | 5.00E+23 | 15.24 | 15.24 | 1.3716 |
| 24 | Ar | 24 | 1.3 | 2.487 | 39.948 | 50.00 | 17.9 | 5.00E+23 | 15.24 | 7.62 | 1.3716 |
| 25 | Ar | 24 | 1.3 | 2.487 | 39.948 | 75.00 | 26.8 | 5.00E+23 | 15.24 | 15.24 | 1.3716 |
| 26 | Ar | 24 | 1.3 | 2.487 | 39.948 | 75.00 | 26.8 | 5.00E+23 | 15.24 | 7.62 | 1.3716 |
| 27 | Ar | 60 | 1.3 | 2.487 | 39.948 | 50.00 | 17.9 | 5.00E+23 | 15.24 | 15.24 | 1.3716 |
| 28 | Ar | 60 | 1.3 | 2.487 | 39.948 | 50.00 | 17.9 | 5.00E+23 | 15.24 | 7.62 | 1.3716 |
| 29 | Ar | 30 | 1.3 | 2.487 | 39.948 | 50.00 | 17.9 | 5.00E+23 | 15.24 | 15.24 | 1.3716 |
| 30 | Ar | 30 | 1.3 | 2.487 | 39.948 | 50.00 | 17.9 | 5.00E+23 | 15.24 | 15.24 | 1.3716 |
| 31 | Ar | 30 | 1.3 | 2.487 | 39.948 | 50.00 | 17.9 | 5.00E+23 | 15.24 | 15.24 | 1.3716 |
| 32 | Xe | 30 | 1.2 | 2.500 | 131.293 | 53.16 | 35.8 | 5.38E+22 | 15.24 | 3.81 | 1.3716 |
| 33 | Xe | 60 | 1.2 | 2.500 | 131.293 | 53.16 | 35.8 | 5.38E+22 | 15.24 | 3.81 | 1.3716 |
| 34 | Xe | 30 | 1.2 | 2.500 | 131.293 | 53.16 | 35.8 | 5.38E+22 | 15.24 | 3.81 | 1.3716 |
| 35 | Xe | 30 | 1.2 | 2.500 | 131.293 | 53.16 | 35.8 | 5.38E+22 | 15.24 | 3.81 | 1.3716 |
| 36 | Xe | 30 | 1.2 | 2.500 | 131.293 | 53.16 | 35.8 | 5.38E+22 | 15.24 | 3.81 | 1.3716 |
| 37 | Ar | 30 | 1.1 | 1.506 | 39.948 | 50.00 | 25.0 | 1.23E+23 | 5 | 5 | 1.3716 |
| 38 | Ar | 30 | 1.3 | 1.274 | 39.948 | 50.00 | 25.0 | 1.23E+23 | 5 | 5 | 1.3716 |
| 39 | Ar | 30 | 1.67 | 0.992 | 39.948 | 50.00 | 25.0 | 1.23E+23 | 5 | 5 | 1.3716 |
| 40 | Ar | 30 | 1.3 | 796.178 | 39.948 | 50.00 | 1.0 | 1.23E+23 | 5 | 5 | 1.3716 |
| 41 | Ar | 30 | 1.3 | 199.045 | 39.948 | 50.00 | 2.0 | 1.23E+23 | 5 | 5 | 1.3716 |
| 42 | Ar | 30 | 1.3 | 31.847 | 39.948 | 50.00 | 5.0 | 1.23E+23 | 5 | 5 | 1.3716 |
| 43 | Ar | 30 | 1.3 | 7.962 | 39.948 | 50.00 | 10.0 | 1.23E+23 | 5 | 5 | 1.3716 |
| 44 | Ar | 30 | 1.3 | 0.201 | 39.948 | 50.00 | 63.0 | 1.23E+23 | 5 | 5 | 1.3716 |
| 45 | Ar | 30 | 1.3 | 0.080 | 39.948 | 50.00 | 100.0 | 1.23E+23 | 5 | 5 | 1.3716 |
| 46 | H | 30 | 1.3 | 0.032 | 1.00794 | 50.00 | 25.0 | 1.23E+23 | 5 | 5 | 1.3716 |
| 47 | ²H | 30 | 1.3 | 0.064 | 2.0141 | 50.00 | 25.0 | 1.23E+23 | 5 | 5 | 1.3716 |
| 48 | Th | 30 | 1.3 | 7.399 | 232.0381 | 50.00 | 25.0 | 1.23E+23 | 5 | 5 | 1.3716 |
| 49 | Ar | 30 | 1.3 | 1.274 | 39.948 | 50.00 | 25.0 | 1.23E+23 | 5 | 5 | 0.3 |
| 50 | Ar | 30 | 1.3 | 1.274 | 39.948 | 50.00 | 25.0 | 1.23E+23 | 5 | 5 | 10 |

### III.A. Physics Description of Implosion

The physical processes can be broken into stages, which we label plasma jet formation, propagation, merging, liner compression, and liner collapse. The discussion will be facilitated by illustrations using case 12.

*Plasma jet formation.* The plasma jets are formed from the ionization and electromagnetic acceleration in a set of railguns or coaxial plasma guns. We assume an initial 3D distribution and launch the jets at the chamber wall radius $r_w$, Fig. 1, which we assume is 1.37 m, the inner radius of the spherical PLX vacuum chamber.

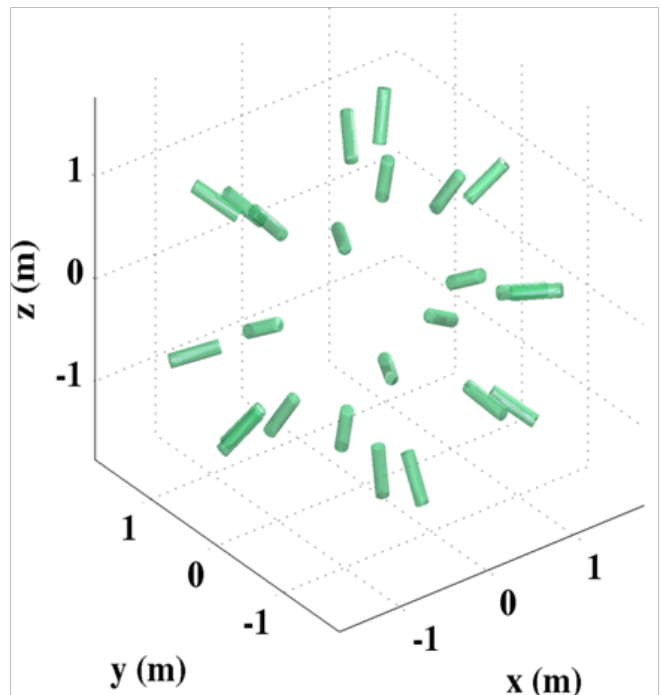

Figure 1. Example of initial plasma jet distribution.

*Jet Propagation.* The jets propagate through a low pressure (~$10^{-6}$ Torr) vacuum until they reach the merging radius $r_M$. This condition occurs approximately when

$$4\pi r_M^2 = N\pi r_{j,M}^2 \tag{4}$$

where $N$ is the number of jets and $r_{j,M}$ is the jet radius at $r_M$. Solving for the jet radius at merging,

$$r_{j,M} = \frac{2r_M}{N^{1/2}} \tag{5}$$

The jet radius is a function of time due to radial expansion into the vacuum. Define the merging time as the point of jet injection at the wall to the merging radius,

$$t_M = \frac{r_W - r_M}{V_{effective}} \tag{6}$$

where $V_{effective}$ is the velocity of the jet plus the longitudinal expansion speed, approximately $2a/(\gamma-1)$ where $a$ is the sound speed. Thus,

$$t_M = \frac{r_W - r_M}{V_j + 2a/(\gamma-1)}$$
or
$$t_M = \frac{r_W - r_M}{a(M_j + 2/(\gamma-1))} \tag{7}$$

Assuming the radial and longitudinal gas expansion rates are the same, the change in jet radius is

$$r_{j,M} - r_{j,0} \equiv \Delta r = \frac{2a}{\gamma-1} t_M \tag{8}$$

Solving for the merge time,

$$t_M = \frac{(r_{j,M} - r_{j,0})(\gamma-1)}{2a} \tag{9}$$

Equating Eqs. 7 and 9,

$$\frac{r_W - r_M}{a(M_j + 2/(\gamma-1))} = \frac{(r_{j,M} - r_{j,0})(\gamma-1)}{2a} \tag{10}$$

Using Eq. 5 and solving for $r_M$,

$$r_M = \frac{r_{j,0}\left(M_j \frac{(\gamma-1)}{2}+1\right)+r_W}{1+\frac{2}{N^{1/2}}\left(M_j \frac{(\gamma-1)}{2}+1\right)} \tag{11}$$

Defining $R_{MW} \equiv \frac{r_M}{r_W}$ as the merging to wall radius ratio and $R_{jW} \equiv \frac{N^{1/2} r_{j,0}}{2 r_W}$ as the square root of the initial jet surface area to chamber surface area ratio, the dimensionless form of Eq. 11 is

$$R_{MW} = \frac{1+\frac{2R_{jW}}{N^{1/2}}\left(M_j \frac{(\gamma-1)}{2}+1\right)}{1+\frac{2}{N^{1/2}}\left(M_j \frac{(\gamma-1)}{2}+1\right)} \tag{12}$$

Equation 12 shows that as $R_{jW} \to 1$ or $N \to \infty$ the jet merging radius approaches the chamber radius. When $M \to \infty$ the equation reduces to $r_M = N^{1/2} r_{j,0}/2$ which is the result for ballistic jet propagation (no expansion). It is important to point out that Eq. 12 gives the maximum merging radius. The expansion rate may be reduced by various effects, in particular, inflight jet cooling or finite background pressure in the chamber. Consequently, the merging radius will tend to be less than the value calculated with the above equation. A lower bound can be determined using the infinite Mach number limit. In other words,

$$r_W R_{MW} > r_M > \frac{N^{1/2} r_{j,0}}{2} \tag{13}$$

accounting for real world effects, such as imperfectly distributed jets, or jets with a radial component of velocity introduced by the acceleration process. For a comparable PLX estimate, we choose Case 7 from Table 1. The chamber diameter is 2.74 m (9 ft), initial jet diameter is 8 cm (~ 3.15 inches), M~25, and the number of jets is 30. With these parameters, the analytical result predicts a merging radius of 0.57 m. By comparison, SPHC predicts some of the jets begin to touch at about 0.5 m (Fig. 2). It should be noted that the actual jet merging radius will vary among the jets and depend on the initial jet configuration.

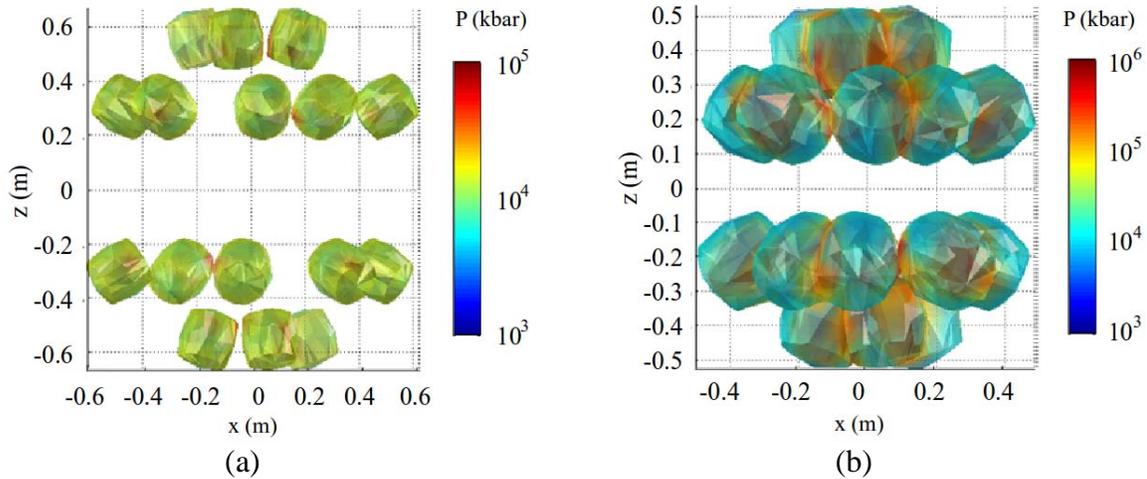

Figure 2. Actual jet merging radius (a) when some jets begin to touch (~0.5 m), and (b) jets in each hemisphere all touching (~0.25 to 0.3 m).

*Liner compression and collapse.* Once the liner has formed, the implosion continues until the inner liner surface reaches either the target boundary or collapses at the origin. Analytical solutions exist for the cavity collapse problem and self-similar converging shocks[35], but there is no solution in general for implosions of liners with fixed boundaries. To a rough approximation, the peak pressure can be estimated to be the incoming liner ram pressure, but the liner ram pressure may amplify considerably from the merge radius down to the target surface or cavity origin.[18] This is further complicated by asymmetries present in the liner caused by the formation by discrete jets.

A qualitative description of the process can be facilitated by a typical time history of peak pressure (Fig. 3) for cases involving 12, 18, 24, and 60 jets (cases 9-11 and 13), respectively. Note that in this figure, peak pressure is the maximum pressure that occurs within the flowfield at a given instance in the simulation, and the location may vary somewhat. For the peak pressure history plot, this was necessary in order to capture the pressure history before void collapse. A sharp, but continuous rise occurs on a time scale of $\sim r_M/V_j$ as the jets coalesce and the liner converges. A spike in the pressure follows, caused by the liner collapse, which launches a radially outward shock.

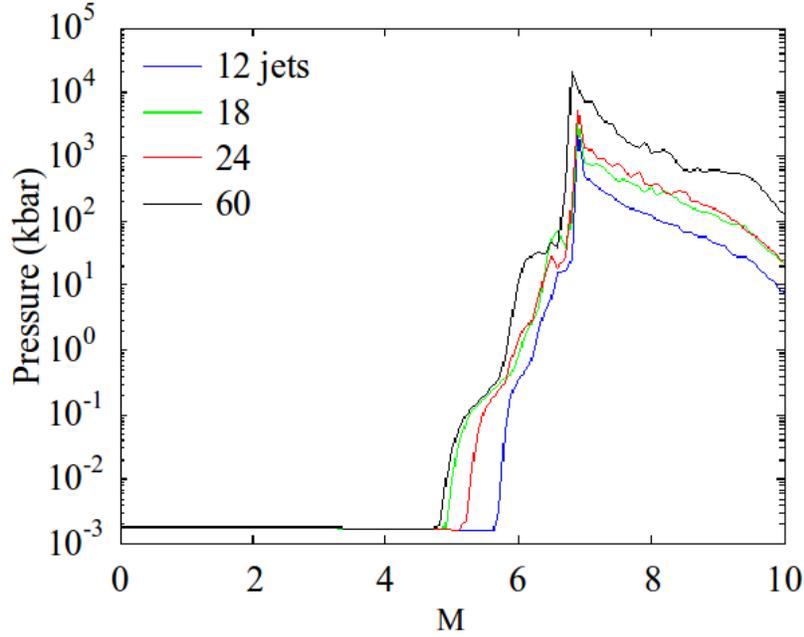

Figure 3. Example of peak pressure history for 12, 18, 24, and 60 jets.

Among all cases investigated, the time to $\sim 1/8$ of the peak is approximated by the time scale $l_j/V_j$, so that the dimensionless stagnation time is

$$\bar{t} \equiv t_{stag} \frac{V_j}{l_j} \tag{14}$$

The actual and scaled stagnation times are shown vs. Mach number in Fig. 4, and the stagnation time is consistent with results predicted in Ref. 25.

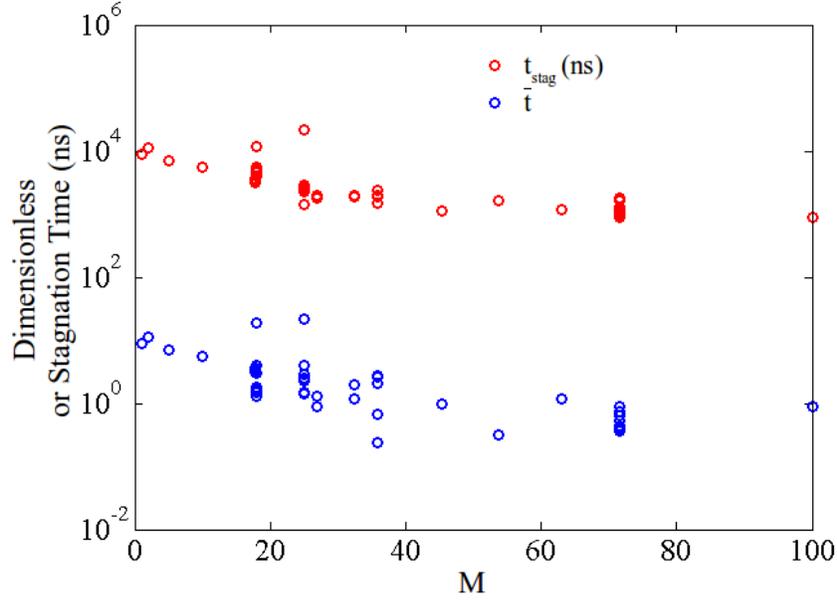

Figure 4. Time from peak to 1/8 of absolute peak pressure vs. Mach number. $\bar{t}$ is nondimensionalized by $l_j/V_j$.

### III.B. Determination and Interpretation of Peak Pressure

Note that the absolute peak pressure in time and space for the simulation occurs at the origin right after void collapse. This peak typically occurs at one or a few particles for just a few time steps near the origin. Since one of the primary objectives of this work is to develop a scaling relation for peak pressure as a function of the multidimensional parameter space defined by the implosion of 3D jets, quantitative determination of peak pressure requires an approach more rigorous than merely selecting the maximum pressure, so the results can be interpreted in a consistent manner. Thus, the peak pressures used in the scaling relation are interpolated precisely at the origin using the numerical output, and are qualitatively consistent with the absolute peak pressures, but typically lower by a factor of ~5.

To determine the peak pressure, the data from each time step were utilized to interpolate the value of pressure at the origin using the summation interpolant, Eq. 2, with the cubic spline being utilized for the smoothing kernel

$$W(R,h) = \frac{3}{2\pi h^3} \begin{cases} \frac{2}{3} - R^2 + \frac{1}{2}R^3 & 0 \leq R < 1 \\ \frac{1}{6}(2-R)^3 & 1 \leq R < 2 \\ 0 & R > 2 \end{cases} \quad (15)$$

where $R$ is given by

$$R = \frac{\left[(x-x')^2 + (y-y')^2 + (z-z')^2\right]^{1/2}}{h} \quad (16)$$

and where x, y, and z are all 0.0 at the origin. The smoothing length $h$ for each particle was calculated with

$$h = \frac{1}{2.197}\left(\frac{m}{\rho}\right)^{1/3} \quad (17)$$

as suggested by Monaghan[23] and each term $W$ in the summation was linearly scaled such that the sum of the weights equaled unity. This approach was tested on randomly scattered points in 3D space in which a hypothetical property '$Q$' was assigned comparable to the pressure field in an imploding liner according to the Gaussian function

$$Q(x,y,z) = 10^{10} e^{-36(x^2+y^2+z^2)} \quad (18)$$

where the parameters were chosen to give peaks and gradients similar to a typical pressure variations found in the simulations of this work. The summation interpolant was then applied to a 2D slice in the z=0 plane to determine the error. For average interparticle spacing of ~5 mm, which is similar to the sph particle spacing at peak compression, the maximum and mean errors were found to be 2.7% and 0.3%, respectively, while the error at the origin was 0.15%. Note that particle spacing can be less than 1 mm in the simulations, but 5 mm gives a worst case scenario in terms of interpolation accuracy.

Radial profiles are of interest to facilitate the interpretation of the peak pressure and hot spot structure produced by plasma liner implosion onto vacuum. The pressure was averaged over the solid angle at fixed radii to produce radial pressure profiles for times of interest, Fig. 5. Case 7 was selected because the jet parameters and total energy are within the range of a 30 jet experiment on the PLX chamber. Upon merging of the jets, the profile is marked by a relatively sharp leading edge which peaks and the decreases gradually with radius towards the outer edge. At 17.5 μs, the void collapses as the leading edge of the liner reaches the center. By 17.9 μs, the pressure profile is relatively flat from the center, extending ~20 cm. The pressure in the center rises ~3 orders of magnitude in the next 1.3 μs, reaching a peak at the center which falls rapidly with radius. The hot spot falls two orders of magnitude within 5 cm, and decreases at a slower rate for the next 20 cm towards the trailing edge of the liner. This hot spot consists of two distinct features, Fig. 6. The central part is a single peak that is fairly symmetric with a radius of ~2 cm, and a lower pressure region surrounding this peak contains very sharp gradients at the outer edge and 'lumps' indicative of the discrete plasma jets. Following the moment pressure reaches a maximum, the hot spot gradually gives way to a smoother pressure profile, while the pressure remains highest at the center.

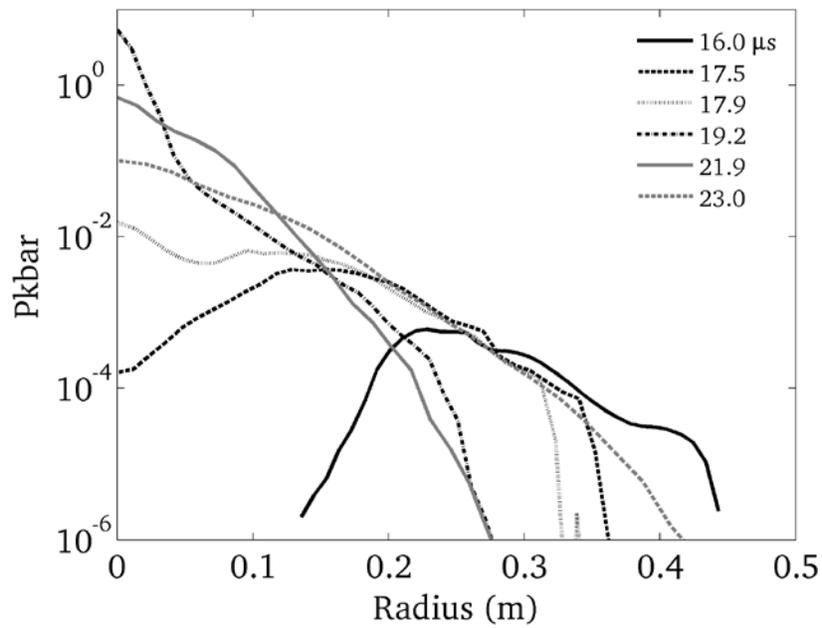

Figure 5. Solid angle averaged pressure vs. radius at fixed times for Case 7.

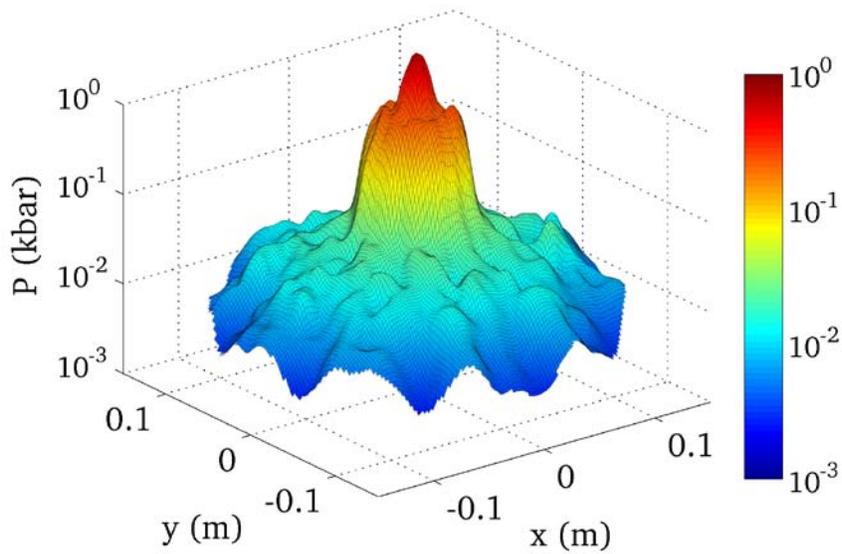

Figure 6. Pressure slice at z=0 plane at 19.2 μs (time of peak pressure) for Case 7.

### III.C. Scaling of Peak Pressure

The objective of this section is to develop a dimensionless pressure which accounts for initial jet conditions. First the scaling relation technique utilized in the derivation is described. This technique is then applied to finding peak pressure as a function of initial plasma jet parameters. Numerical results are presented alongside the scaling results. We relate the scaling relation to uniform imploding shells, and finally give an example of how the relation can be applied for experimental planning purposes.

#### III.C.1. Scaling Relation Techniques

Dimensional analysis is a means for reducing the number of variables that affect some physical phenomenon of interest, and basic techniques are covered in many textbooks (e.g. Ref. 35). It should also be mentioned that a feature article on the subject recently appeared in *Physics Today*.[38] There seems to be no preferred way to develop scaling relations in fusion energy science, and frequently, no specific scaling technique is discussed. For example, Levedahl and Lindl[39] derived a criterion for ignition threshold as a function of the implosion velocity and compressibility of an imploding fuel mass. They ran simulations with LASNEX[40] with an initial a capsule and radiation drive profile, and assumed a homogeneous scaling factor for resizing of the velocity and spatial scale in the simulation. The procedure for finding the scaling relation was not provided. Likewise, the specific technique for developing the scaling relations of Kemp et al.[20] was not discussed. Herrmann, Tabak, and Lindl[21] based their parameterization on the earlier work of Rosen et al.[41], and again no formal approach is offered in the original source. An exception to this observation is given by Luce et al.[42] in which the authors discuss the Buckingham Pi theorem and scale invariance in application to magnetic fusion experiments. Merits and disadvantages of both approaches are given, with no clear preferred method suggested. Our approach for developing the scaling relations was based on Ipsen's method[43], which is a step-by-step approach that obtains all of the dimensionless variables at once. We have found this method to be a very fast and convenient means of determining relevant dimensionless parameters.

### III.C.2.    Derivation of Peak Pressure Scaling Relation

For the ideal hydrodynamic implosion of plasma jets studied in this paper, peak stagnation pressure is some unknown function

$$P_{peak} = f(r_j, l_j, r_w, N, \rho, V_j, (\gamma RT)_j, \gamma) \tag{19}$$

The basic approach in Ipsen's method is to successively divide through by variables until all the remaining terms are dimensionless. The number of dimensionless terms remaining will typically be the number of variables minus the number of dimensions. In our case that results in 6 dimensionless parameters, such as some function of the form

$$\frac{P_{peak}}{\rho V^2} = f\left(\frac{l_j}{r_j}, \frac{r_w}{r_j}, N, M_j, \gamma\right) \tag{20}$$

The method yielded a fairly typical nondimensional pressure and 5 dimensionless parameters (initial jet length to radius ratio, chamber wall to initial jet radius ratio, number of jets, initial Mach number, and specific heat ratio). Even though the parameters have been reduced from a 9D to a 6D parameter space, it is still difficult to present data with so many variables. Ipsen's method required just a few minutes of work to arrive at Eq. 20, and allowed a starting point from which to simplify the relationship to something more convenient. At this point, any further simplifications are ad hoc. With some trial and error, it was discovered that within about a factor of three (except for special cases to be discussed below), the following scaling relation held:

$$\overline{P} \equiv P_{peak} \frac{2R_w}{\rho V^2} = f(M_j) \tag{21}$$

where $R_w$ is given by

$$R_w = \frac{4r_w^2}{Nr_j^2} \tag{22}$$

and

$$f(M_j) = 5M^2 \tag{23}$$

The scaled results are plotted in Fig. 7 Dimensionless pressure $\overline{P}$ increases monotonically with M. Actual peak pressure is plotted on the secondary y-axis for comparison. It should be noted that the actual pressure may vary by up to 5 orders of magnitude, while the dimensionless values varied within a single order of magnitude. This indicates that the above expression is a successful model for ideal hydrodynamic scaling.

### III.C.3. Results and Analysis

At fixed *M*, Eq. 21 shows that peak pressure increases linearly with density and number of jets, quadratically with jet radius and initial velocity, and inversely with the square of the wall radius. Qualitatively, these results should be of no surprise. The pressure behind a normal shock can be shown to be

$$P_2 = P_1\left(1 - \frac{2\gamma}{\gamma-1}\right) + \frac{2}{\gamma+1}\rho_1 V_1^2 \tag{24}$$

For the case when $\rho_1 V_1^2 \gg P_1$ (i.e. high *M*), the shock pressure scales with the dynamic pressure, as in Eq. 21. The first term tends to reduce the shocked pressure $P_2$ when $\rho_1 V_1^2 \sim P_1$. This can lower peak pressure in the imploding liner in several ways, such as an initially low *M*, high $r_w$, or low *N*. When the $r_w$ is high, the merge radius will be larger, and a higher compressional work is required before the liner reaches the target or cavity origin, which raises the liner temperature and lowers the liner Mach number. A low *N* increases the angular separation between jets, which increases the strength of the shock between jets during merging, thereby increasing the temperature of the newly formed liner at the merge radius, and depressing the liner Mach number.

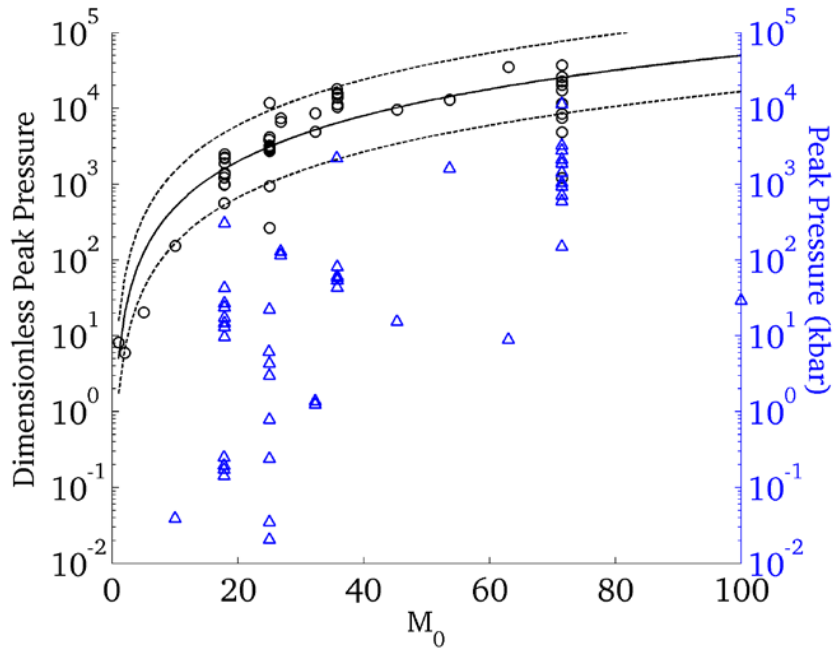

Figure 7.  Dimensionless peak pressure vs. Mach number (black circles). Scaling relation, Eq. 21, shown as solid black line, with 3× and 1/3× this expression shown to bound most of the data. Actual peak pressure (kbar) is shown for reference (blue triangles on the right hand y-axis).

The full black line in Fig. 7 is the $5M^2$ scaling relation. The dashed lines above and below the scaling relation function represent 3 times and 1/3 of the scaling function values, respectively, showing that the vast majority of the 50 runs fall well within this range.

The 3D simulation result of Kim *et al.*[16] appear consistent with our scaling relation. Using the initial plasma jet conditions given in Sec. II of Ref. 16, our scaling relation gives a peak pressure of 4.72 kB compared to their reported result of 6.4 kB

### III.C.4. Departures from the Scaling Relation

There are significant departures from the curve in Fig. 7. Note the few circles significantly lower than the trend line at $M$=71.6 (runs 19, 20, 21 in Table 1). These were cases in which the arrival time of the jets to the merging radius was varied by up to 100 or 1000 ns, to mimic firing jitter in the plasma guns, which can lower the peak pressure by a factor of 5 or 10. A 10 ns jitter case was also run, but it had a negligible effect.

Runs 38, 49, and 50 can be compared for the effects of chamber radius at fixed parameters for $r_w$ = 1.3716, 0.3, and 10 m, respectively. For relatively small $r_w$ (runs 38 and 49), including the PLX chamber radius, the effect was negligible. A large chamber radius appears to have a pronounced and deleterious effect on peak pressure, as observed by the lowest point at $M$=25 in Fig. 7, run 50 in Table 1. The dimensionless peak pressure in this case was 64.7, compared with 616 and 816 for runs 49 and 38, respectively. The radial and longitudinal expansion prior to jet merging could play a significant role as discussed in the previous section. In runs 38, 49, and 50, both the initial jet diameter and length were 5 cm. At the merging time the growth in both length scales can be determined with Eq. 8, which predicts both a linear and radial growth in the jet by a factor of 2.1, 6.6, and 43 times prior to merge for the chamber radii of 0.3, 1.3716, and 10 m, respectively. Alternatively, this can be clarified by comparing the merge time with a characteristic expansion time $t_{j,exp}$, here defined as the time required for a jet to expand to twice the initial jet length or radius,

$$t_{j,\exp} = \frac{r_{j,0}(\gamma-1)}{2a} \tag{25}$$

For these three cases, the merge times are approximately 1, 6, and $42 \times t_{j,exp}$. Thus, when $t_M \gg t_{j,exp}$, the peak pressure can be expected to be significantly lower. The qualitative observations are to be expected, since longitudinal expansion will lower the effective dynamic pressure at the cavity collapse, while radial expansion may increase the heating via increased strength in the shocks or other compression waves at jet interfaces during the merging process.

Some mention should also be made of runs 40 through 43 in Table 1, which have $M<10$. Only a few $M<10$ cases were run, so there is not enough data to justify labeling these two points as 'departure'. The sharp decrease in peak pressure as $M \to 1.0$ is consistent with normal shock theory and with other pressure scaling relations for implosions,[21] and shows that formation of a high $M$ liner is critical to achieving high pressures.

### III.C.5. Scaling Parameter in Terms of Spherically Symmetric Liners

The scaling relation derived above (Eq. 17) can be related to spherically symmetric liners (rather than initially discrete jets) with the assumption of large $r_W$ to liner thickness ratio, if the initial jets are replaced by a uniform shell of the same total mass, with a thickness equal to $l_j$. To see this, first, the total liner mass is given by

$$m = \rho \pi r_j^2 l_j N \tag{26}$$

In terms of $m$, Eq. 21 in terms of the total liner mass is

$$\overline{P} \equiv P_{peak} \frac{2}{V^2}\left(\frac{4\pi r_w^2 l_j}{m}\right) = f(M_j) \qquad (27)$$

The density of a spherical shell of the same $m$, inner radius $r_w$, and thickness equal to $l_j$ is given by

$$\rho_{shell} = \frac{m}{\{4/3\pi[(r_w+l_j)^3 - r_w^3]\}} \qquad (28)$$

For $l_j \ll r_w$,

$$\rho_{shell} \approx \frac{m}{(4\pi r_w^2 l_j)} \qquad (29)$$

Noting that the RHS of Eq. 29 is identical to the reciprocal of the term in parentheses in Eq. 27, so we have

$$\overline{P} \equiv P_{peak} \frac{2}{\rho_{shell} V^2} = f(M_j) \qquad (30)$$

Thus, the nondimensional parameter $\overline{P}$ is equivalent to the peak pressure nondimensionalized against the dynamic implosion pressure of a spherically symmetric plasma liner of the same mass and radial implosion velocity, and can be applied to 1D simulations. Also, from Eq. 27, it is seen that at fixed $M$, the pressure scales linearly with the initial radial kinetic energy and inversely with the initial jet length. Peak pressure decreases with $r_w^2$ because the liner will lose kinetic energy to thermal energy due to increased compressional work done by the liner.

The scaling relation given by Eq. 30 was applied to the 1D simulations of Awe et al.,[12] Davis et al.,[13] and Kim et al.,[15] and plotted in Fig. 8. The data from Awe et al. were generated using the 1D radiation-hydrodynamic code Raven[44] in which thermal conduction and radiation were included, and the equations of motion were closed with a constant γ=5/3 ideal gas. Davis et al.

utilized Helios[45] with thermal conduction, radiation, and a tabular EOS model generated by the Propaceos[45] code. Finally, Kim *et al.* produced their results with the FronTier code[46] and included ionization in their EOS model as well as thermal conduction. The peak pressures observed in the 1D results from all authors either are consistent with the scaling relation or are considerably higher (Fig. 8). One of the main observations in Awe *et al.* was that radiation plays a crucial role in enabling much higher pressures, as is apparent in Fig. 8 below. Specific reasons for the departures due to thermal transport and radiation are beyond the scope of this work but will be addressed, along with the effects of ionization and electronic excitation, in a future paper utilizing the 3D radiation hydrodynamic code SPHC.

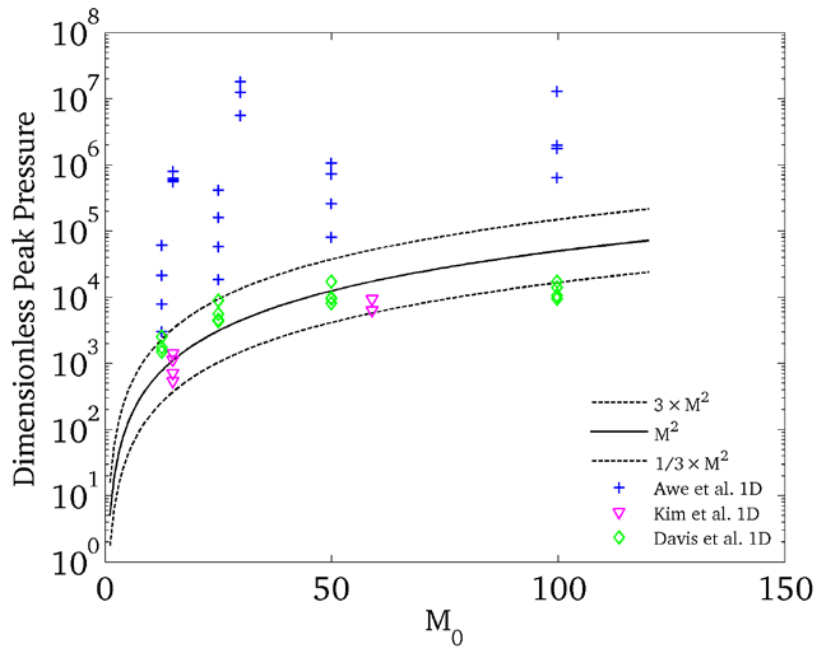

Figure 8. One dimensional scaled peak pressure versus Mach number from Awe, *et al.*,[12] Davis *et al.*,[13] and Kim *et al.*[15] The scaling relation, Eq. 30, shown as solid black line, with 3× and 1/3× of this expression shown for reference as in Figure 7

### III.C.6. Application of the Scaling Relation for Experimental Design

As a final note on dimensionless scaling, the requirements for a 1 Mbar experiment can be estimated using Eq. 21. Choosing $\rho$=0.67 kg/m$^3$, V= 75 km/s, $r_j$=8 cm, $r_w$=1.3716 m, N=60 jets, and a jet Mach number of 15, the scaling relation predicts a peak pressure of ~1 Mbar. A jet length of 10 cm would allow the imploded plasma to maintain pressure within an order of magnitude of the peak for about 1 µs. It must be emphasized again that the inclusion of radiation transport can possibly give higher peak pressure due to more effective compression. This will be investigated in future 3D SPHC simulations.

In summary, to achieve higher peak pressures, one must increase *M*, increase the total initial kinetic energy, decrease the initial thickness of the liner, or a combination of all three. With an understanding of the relationship between jet parameters and peak pressure based on the dimensionless approach above, one can then determine the absolute scale required based on the total liner kinetic energy.

### III.D. Dependence of Peak Pressure on Total Liner Kinetic Energy

While the previous section demonstrates the scale invariance of the dimensionless peak pressure versus *M*, Figure 7 shows the roughly linear dependence of absolute peak pressure on kinetic energy. We include the 1D simulations from Awe *et al*,[12] Davis *et al*.,[13] and Kim *et al*.[15] to illustrate non-ideal effects.

First we comment on the spread in the 3D ideal gas SPHC data at ~30 kJ. The four lowest data points are runs 40 through 43, having *M*<10. The three highest data points in this group are runs 44, 45 and 49. Runs 44 and 45 have high values of *M* = 63 and *M* = 100, respectively. Run 49 has a small jet injection radius of 0.3 m, which gives the jets less time to expand, lowers the liner compressional work, and lowers the effective inter-jet merge angle.

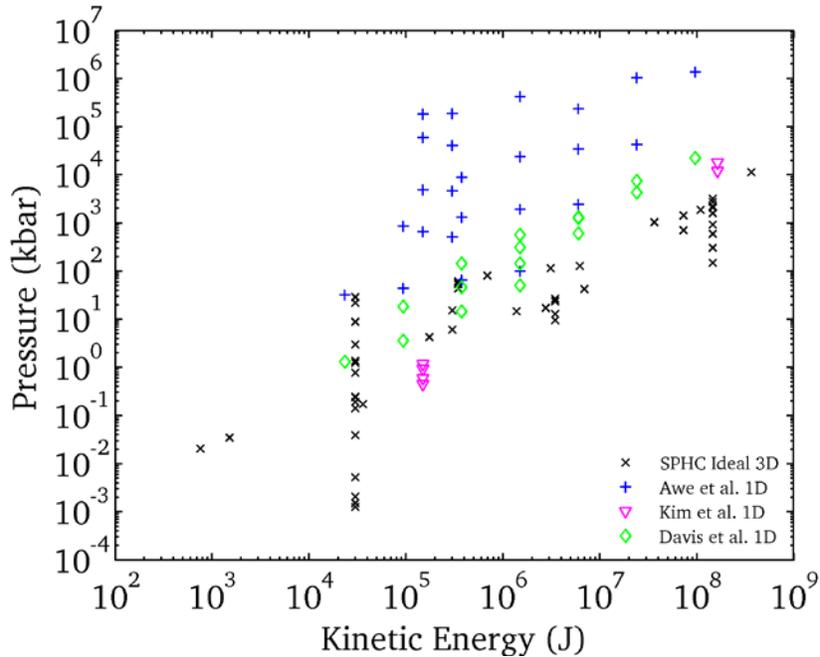

Figure 9. Peak pressure in kbar vs. kinetic energy of the liner.

Next note the trend in the 1D simulations which include at least some combination of radiation, thermal conduction, and ionization. All of these data give pressures that are comparable to or greater than the ideal 3D SPHC data, or give higher peak pressures, Figure 9. The results from Awe *et al*. give peak pressures that spread considerably more compared with Davis *et al*. and Kim *et al*. These simulations did not include the effects of internal energy states or ionization and utilized a constant gamma ideal gas law. Davis *et al*. utilize the same code but with a tabular EOS model, showing the importance of including ionization. While further modeling needs to be performed in 3D, the comparisons among these data suggest that radiative and thermal transport will facilitate much higher compression, while this affect will be compromised somewhat by ionization processes in the liner. The ideal 3D SPHC data appear tentatively to give a lower limit on the expected peak conditions.

## IV. Conclusions

The Plasma Liner Experiment (PLX) was designed to explore and demonstrate the feasibility of forming imploding spherical "plasma liners" that can reach peak pressures ~0.1-1 Mbar upon stagnation by the merging of 30 plasma jets in a spherical configuration. PLX is motivated by the possibility of plasma liner driven MIF and by a new approach to reaching the HED regime in the laboratory. To assist in the planning and assessment of the requirements of the jets for the experiments, a series of 3D simulations were performed using SPH.

Scaling relations were developed using the numerical results to estimate the merging radius, stagnation time at peak compression, and peak pressure. The merging radius scales as $N^{1/2} r_{j,0}/2$ in the ballistic limit, but is otherwise higher due to the thermal expansion of the jets. At peak compression the stagnation time is approximately $l_j/V_j,$ which is consistent with results predicted in Ref. 34. Peak pressure increases linearly with density and number of jets, quadratically with jet radius and initial velocity, and inversely with the initial jet length or square of the wall radius. Within about a factor of three (except for special cases), the scaling relation for peak pressure was found to monotonically increase with initial jet Mach number. The departure from the scaling relation was within a factor of 3, including with results from a separate hydro code Frontier, which was astonishing considering the large number of initial jet parameters. Departures outside a factor of 3 were due to aspects not accounted for in the scaling relation, including small perturbations in the initial jet timing to mimic experimental plasma gun firing jitter, and large chamber radii in which the jet expansion time was much less than the jet merging time. Both effects were found to reduce the peak pressure by an order of magnitude. 1D simulations from Awe et al.,[12] Davis *et al.*,[13] and Kim *et al.*[15] were either in agreement or exceeded the predicted peak pressure, illustrating the importance of including radiation, thermal transport, and ionization. The departures from the trend enable the isolation of the effects of a single parameter or attribute on the peak pressure. Including radiation in future 3D studies is clearly needed, and the 1D results thus far suggest that by including radiation, the liner is able to achieve greater convergence and thus higher peak pressure. Future work will include effects gradients in the initial jets, other jet shapes, ionization, and thermal and radiative transport in 3D.

**Acknowledgments**

This work was supported in part by the U.S. Department of Energy under grants/contracts DE-SC0003560, DE-AC52-06NA25396, DE-FG02-05ER54810 and NSF-1004330. The authors thank Francis Thio and Doug Witherspoon for many helpful discussions.